\documentclass[conference]{IEEEtran}
\usepackage[ruled,linesnumbered,vlined]{algorithm2e}
\usepackage{cite,graphics,graphicx}
\usepackage{multirow,makecell}
\usepackage{amssymb}
\usepackage{url}
\usepackage{amsmath}
\usepackage[misc]{ifsym}

\IEEEoverridecommandlockouts
\usepackage{amssymb,amsfonts}
\usepackage[hyphenbreaks]{breakurl}
\usepackage{algorithmic}
\usepackage{flushend}
\usepackage{graphicx}
\usepackage{textcomp}
\usepackage{xcolor}
\usepackage{booktabs}
\usepackage{listings}
\usepackage{bbding}
\usepackage{float}
\def\BibTeX{{\rm B\kern-.05em{\sc i\kern-.025em b}\kern-.08em
    T\kern-.1667em\lower.7ex\hbox{E}\kern-.125emX}}
\begin{document}

\title{NUMA-aware FFT-based Convolution on ARMv8 Many-core CPUs\thanks{This research work was supported by the National Natural Science Foundation of China under grant No. 62002365, and the National Key Research and Development Program of China under grant No. 2018YFB0204301.}
}

\author{\IEEEauthorblockN{Xiandong Huang\IEEEauthorrefmark{1}\IEEEauthorrefmark{2}, Qinglin Wang\Envelope\IEEEauthorrefmark{1}\IEEEauthorrefmark{2}, Shuyu Lu\IEEEauthorrefmark{3}, Ruochen Hao\IEEEauthorrefmark{1}\IEEEauthorrefmark{2}, Songzhu Mei\IEEEauthorrefmark{1}\IEEEauthorrefmark{2}, Jie Liu\IEEEauthorrefmark{1}\IEEEauthorrefmark{2}}
	\IEEEauthorblockA{\IEEEauthorrefmark{1}Science and Technology on Parallel and Distributed Processing Laboratory\\
		National University of Defense Technology, Changsha 410073, China}
	\IEEEauthorblockA{\IEEEauthorrefmark{2}School of Computer Science, National University of Defense Technology, Changsha 410073, China}
	\IEEEauthorblockA{\IEEEauthorrefmark{3}Department of Biomedical Informatics, University of Pittsburgh, Pittsburgh, USA}
	\IEEEauthorblockA{Email: huangxiandong@nudt.edu.cn, wangqinglin.thu@gmail.com}}

\maketitle
\begin{abstract}
Convolutional Neural Networks (CNNs), one of the most representative algorithms of deep learning, are widely used in various artificial intelligence applications. Convolution operations often take most of the computational overhead of CNNs. The FFT-based algorithm can improve the efficiency of convolution by reducing its algorithm complexity, there are a lot of works about the high-performance implementation of FFT-based convolution on many-core CPUs. However, there is no optimization for the non-uniform memory access (NUMA) characteristics in many-core CPUs. In this paper, we present a NUMA-aware FFT-based convolution implementation on ARMv8 many-core CPUs with NUMA architectures. The implementation can reduce a number of remote memory access through the data reordering of FFT transformations and the three-level parallelization of the complex matrix multiplication. The experiment results on a ARMv8 many-core CPU with NUMA architectures demonstrate that our NUMA-aware implementation has much better performance than the state-of-the-art work in most cases. 
\end{abstract}


\section{Introduction}
Convolutional Neural Networks (CNNs) have been widely adopted in various fields of artificial intelligence \cite{kurth2018exascale,george2018deep,8735483,litjens2017a}. For instance, the scientific community applies CNNs into scientific analysis and discovery \cite{kurth2018exascale, george2018deep}, and many works focus on analyzing medical images using CNNs \cite{litjens2017a}. However, it's very time-consuming to train CNNs. Convolution operations often take more than 80\% of the computational overhead of CNNs. Therefore, high-performance implementation of convolution operations is particularly important for improving the computational efficiency of CNNs.

The popular algorithms for implementing convolution operations are matrix multiplication-based, Winograd-based, Fast Fourier Transform (FFT)-based and direct algorithms \cite{9183396, jia2014caffe, ijcnnwang2019, zhang2018high, lavin2016fast, winograd2020wang, wang2020optimizing}. The FFT-based algorithm converts convolutions into complex matrix multiplication by means of fast Fourier transform, and includes four stages: input transformation, kernel transformation, complex matrix multiplication, and output transformation. Compared with other three algorithms, the FFT-based convolution has lower arithmetic complexity while the accuracy loss can be ignored. Thus, there is a lot of work about studying high-performance implementation of FFT-based convolution algorithm on different platforms. Mathieu and Vasilache et al. \cite{VasilacheJMCPL14, mathieu2014fast} first proposed how to implement convolutions by the FFT-based algorithm on modern GPUs, and introduced different FFT-based convolution implementations on GPUs. Zlateski et al. \cite{zlateski2016znn, zlateski2016znninference, zlateski2018fft} mainly put their optimizing efforts into the implementations of FFT–based convolutions on Intel multi– and many–core CPUs. Lin et al.\cite{2019lin} presented a decomposition strategy to optimize FFT convolution on GPUs. Wang et al. \cite{wang2020optimizing} proposed a parallel FFT-based convolution implementation on ARMv8 CPUs, which does not depend on any other computing libraries. However, none of the work above involves the utilization of non-uniform memory access (NUMA) characteristics in modern many-core CPUs. 

The state-of-the-art many-core system designs often use NUMA techniques to scale the number of cores and the memory bandwidth. Some vendors have built processors with tens of cores based on the NUMA architecture, such as AMD EPYC series \cite{amdEPYC3} and Phytium FT-2000plus \cite{FT2000}. Further, multiple processors can be organized into a multi-socket computing node in a NUMA manner. This paper mainly focuses on the former, many-core processors with NUMA architecture. In a many-core CPU with NUMA architecture, a core can directly access the local memory of the NUMA node it belongs to, and access the remote memory attached to the other NUMA nodes through the network-on-chip. Thus, when the core and the memory are located in different NUMA nodes, the memory access latency will increase significantly. In other words, parallel implementations without considering NUMA characteristics may get very bad performance on such many-core CPUs.

In this paper, we propose a NUMA-aware FFT-based convolution algorithm named nFFT, targeted at the convolution optimization on ARMv8 many-core CPUs with NUMA architectures. The complex matrix multiplication occupies most of the total computing overhead of the FFT-based convolution algorithm. Therefore, the design philosophy of our algorithm is improving the performance of the complex matrix multiplication as much as possible without significantly increasing the overhead of the left stages. In nFFT, we design the three-level parallelization of the complex matrix multiplication, and reorder the results of input and kernel transformations to minimize the number of remote memory access in the complex matrix multiplication. We benchmark 17 convolution configurations from three famous CNN architectures Alexnet, VGG and Resnet on a ARMv8-based Phytium FT-2000plus 64-core  processor. Compared with the prior implementation \cite{wang2020optimizing} on ARMv8 architectures, our nFFT achieves a speedup of 1.01 - 1.84 times in most cases. In order to understand the improvement, we also make a level-2 cache miss rate comparation between nFFT and the prior.

The structure of this paper is as follows. Section~\ref{background} introduces FFT-based convolution algorithm and the architecture of ARMv8-based Phytium FT-2000plus processor. Section~\ref{analysis} analyzes the influences of NUMA architecture on the performance of FFT-based convolution algorithm. Section~\ref{algimpl} describes our proposed NUMA-aware FFT-based convolution algorithm and implementation on ARMv8 many-core CPUs in detail. The performance results and analysis are placed in Section~\ref{per_result}. Finally, Section~\ref{conclusion} concludes this paper and gives our future work.

\section{BACKGROUND}
\label{background}

\subsection{FFT-based Convolution}
Convolution operation involves three variables: input feature maps ($I$), kernels ($K$) and output feature maps ($O$). In BCHW (batch, channel, height, width) layout, these variables can be written as  $I[B][C][H_i][W_i]$, $K[C'][C][H_k][W_k]$ and $O[B][C'][H_o][W_o]$, where $C'$ and $C$ are the number of output and input channels, $B$ stands for the batch size, and $H_{i/o/k}$ and $W_{i/o/k}$ represent the spatial dimension sizes. According to these data formats, convolution operation in CNNs can be obtained as:
\begin{eqnarray}\label{eq.conv1}
\begin{aligned}
O_{b, c', h_o, w_o} = \sum_{c=0}^{C-1}{\sum_{h_k=0}^{H_k-1}{\sum_{w_k=0}^{W_k-1}}} (I_{b, c, h_o + h_k, w_o + w_k} \\ \times  K_{c', c, h_k, w_k}),
\end{aligned}
\end{eqnarray}
where $0\leq b<B$, $0\leq c'<C'$, $0\leq h_o<H_o$, $0\leq w_o<W_o$, $0\leq c<C$, $0\leq h_k<H_k$, $0\leq w_k<W_k$.

The FFT-based convolution algorithm consists of four stages: the FFT transformation of input feature maps and kernels, the complex matrix multiplication, and the inverse FFT (IFFT) transformation of output feature maps. The equation \ref{eq.conv2} describes these four stages above:
\begin{eqnarray}\label{eq.conv2}
\begin{aligned}
{O_{b,c'}} = {F^{-1}(\sum\limits_{c \in C} F({I_{b,c}}) \cdot F^*({K_{c',c}})} ),
\end{aligned}
\end{eqnarray}
where $F$ and $F^{-1}$ are 2D FFT and IFFT respectively, $\cdot$ represents element-wise multiplication which would be converted to the classical complex matrix multiplication. As $H_{k}$ and $W_{k}$ are often much smaller than $H_{i/o}$ and $W_{i/o}$, the tiling approach is used to minimize the overhead of padding. If there is no tiling approach, the kernel needs to be expanded to the size of the input feature map, which requires lots of overhead of padding. We label the coordinates of a tile in the tiling as $(\alpha, \beta)$. So, the FFT-based convolution algorithm can be expressed as:
\begin{eqnarray}\label{eq.conv3}
\begin{aligned}
{O_{b,c',\alpha,\beta}} =  {F^{-1}(\sum\limits_{c \in C} {F({I_{b,c,\alpha,\beta}}) \cdot F^*({K_{c',c}}})} ).
\end{aligned}
\end{eqnarray}

Each element of the element-wise multiplication is labeled as $(\varphi, \gamma)$. There are a total of $P$ elements in the element-wise multiplication. The third stage of FFT-based convolution algorithm is actually a batched complex matrix multiplication as follows:
\begin{eqnarray}\label{eq4.cmm}
\begin{aligned}
{Z^{(\varphi ,\gamma )}} = {G^{(\varphi ,\gamma )}}{D^{(\varphi ,\gamma )}},
\end{aligned}
\end{eqnarray}
Where $D_{c,b,\alpha ,\beta }^{(\varphi ,\gamma )} = F{({I_{b,c,\alpha ,\beta }})^{(\varphi ,\gamma )}}$ and $G_{c',c}^{(\varphi ,\gamma )} = F^*{({K_{c',c}})^{(\varphi ,\gamma )}}$. $Z$, $G$, and $D$ are the matrices of $ P \times M \times C'$, $ P \times M \times C$ and $ P \times C \times C'$ sizes, where $M= B \times \rm X \times \Delta$, and $\rm X \times \Delta$ represents the number of tiles in each feature map.

The native implementation and the optimization of FFT-based convolution algorithm without considering NUMA architecture on ARMv8 many-core CPUs can refer to \cite{wang2020optimizing}.

\subsection{Architecture of Phytium FT-2000plus}

\begin{figure*}[htb!]
	\centering
	\includegraphics[width=17cm, height=6.0cm]{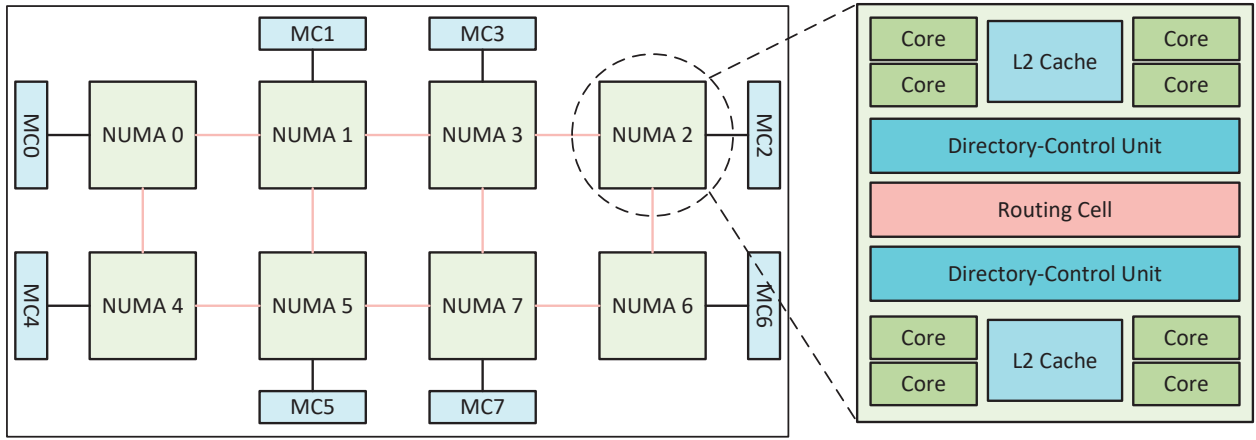}
	\caption{Architecture of Phytium FT-2000plus.}
	\label{fig0:ft2000:arch}
\end{figure*}

Fig.~\ref{fig0:ft2000:arch} shows the architecture of Phytium FT-2000plus, which is the Phytium's second-generation many-core architecture and code-named Mars II \cite{PhyMars2}. It consists of eight NUMA nodes, each of which directly connects a local Memory Controller (MC). All NUMA nodes are connected together by a configurable on-chip network, and each NUMA node can access a remote memory controller through the on-chip network. The memory access bandwidth evaluation \cite{yu2020optimizing} shows that a NUMA node can access local memory at the highest memory bandwidth while the bandwidth of accessing remote memory drops sharply. Thus, it's necessary for the performance optimization to minimize the number of remote memory access. 

Each NUMA node includes eight ARMv8-based cores \cite{Phyxiaomi}, two Directory Control Units (DCU) and one routing cell. The working frequency of these cores is 2.3 GHz. Each core features a super-scalar out-of-order pipeline, and can issue up to four instructions per cycle. The vector units in each core can deal with 4 single precision operations at a time. Each core has a 32 KB private L1 instruction cache and a 32 KB private L1 data cache. Four cores share a 2 MB inclusive L2 cache. All the caches of eight NUMA nodes are kept coherent by 16 distributed DCUs. 

\section{Analysis}
\label{analysis}
As mentioned in the previous section, FFT-based convolution algorithm takes two tensors ($I$ and $K$) as input and one tensor ($O$) as output, and includes four stages: the transformations of input feature maps and kernels, the complex matrix multiplication, and the inverse transformation of output feature maps. The prior optimization implementation of FFT-based convolution (named wFFT) \cite{wang2020optimizing} also involves three other tensors ($D$, $G$ and $Z$), which are used to store the result of the first three stages, respectively. In order to obtain stable performance on many-core systems with NUMA architecture, the Linux command \textit{numactl --interleave=all} is often executed to evenly distribute the data of each tensor on all NUMA nodes. 

\begin{figure*}[htb!]
	\centering
	\includegraphics[width=17cm, height=6.5cm]{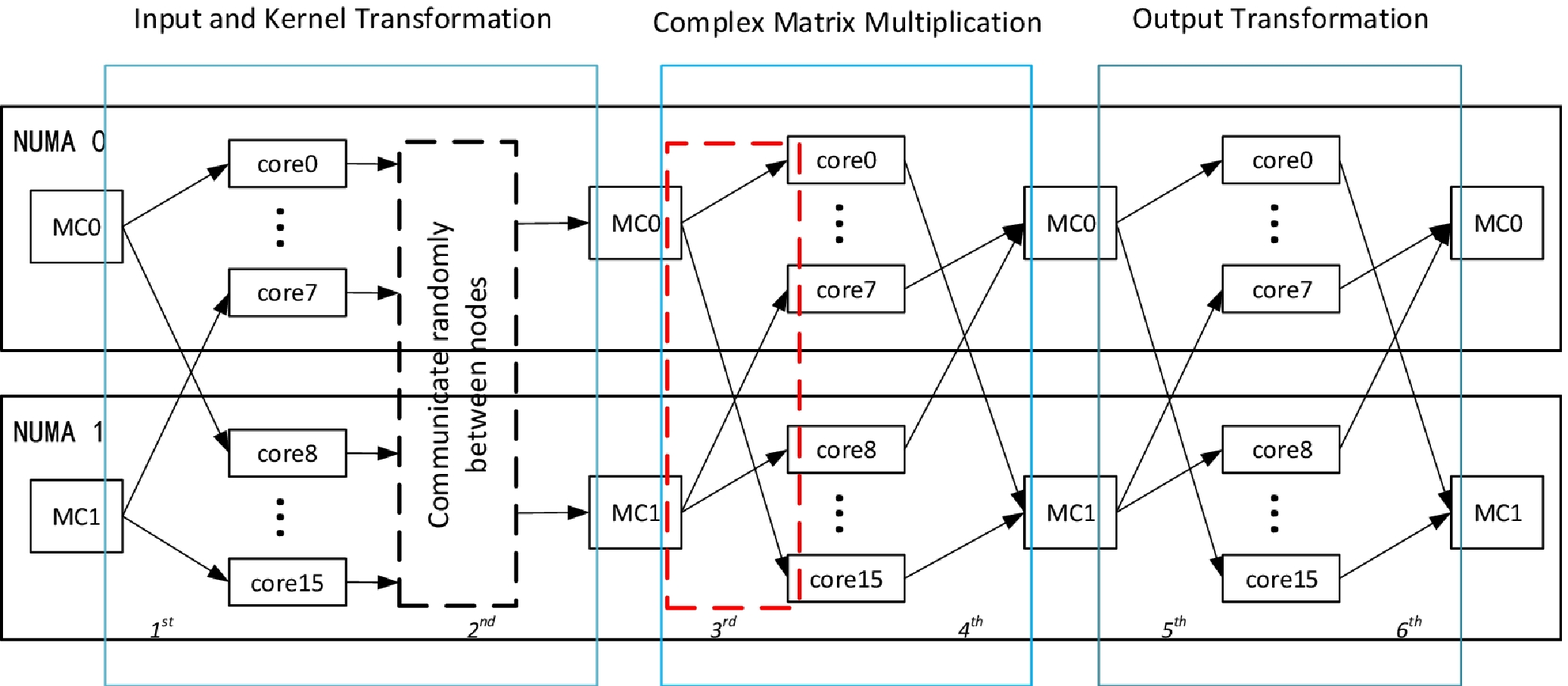}
	\caption{NUMA communication in wFFT.}
	\label{fig1:comm:prior}
\end{figure*}

Fig.~\ref{fig1:comm:prior} illustrates the communication between NUMA nodes in wFFT. We can find that all four stages load input tensors from all NUMA nodes and store output tensors back to all NUMA nodes. There may be a large number of remote memory accesses in each stage, drastically increasing the total overhead. In this paper, the input and output tensors ($I$, $K$ and $O$) of convolution operations are also uniformly stored on all NUMA nodes, and it is not cost-effective to find the node where each element of these tenors is located. So the remote memory access in the fetching ($1^{st}$ in Fig.~\ref{fig1:comm:prior}) of the first two stages and the writing ($6^{th}$ in Fig.~\ref{fig1:comm:prior}) of the last one stage can not be eliminated or reduced. And the complex matrix multiplication often takes most of the total overhead of the FFT-based convolution algorithm. Thus, our goal is to avoid or eliminate remote memory access in the complex matrix multiplication so as to maximize its performance, while the overhead of the left three stages is not significantly increased.

\section{Algorithm and Implementation}
\label{algimpl}

\begin{figure*}[htb!]
	\centering
	\includegraphics[width=17cm, height=6.5cm]{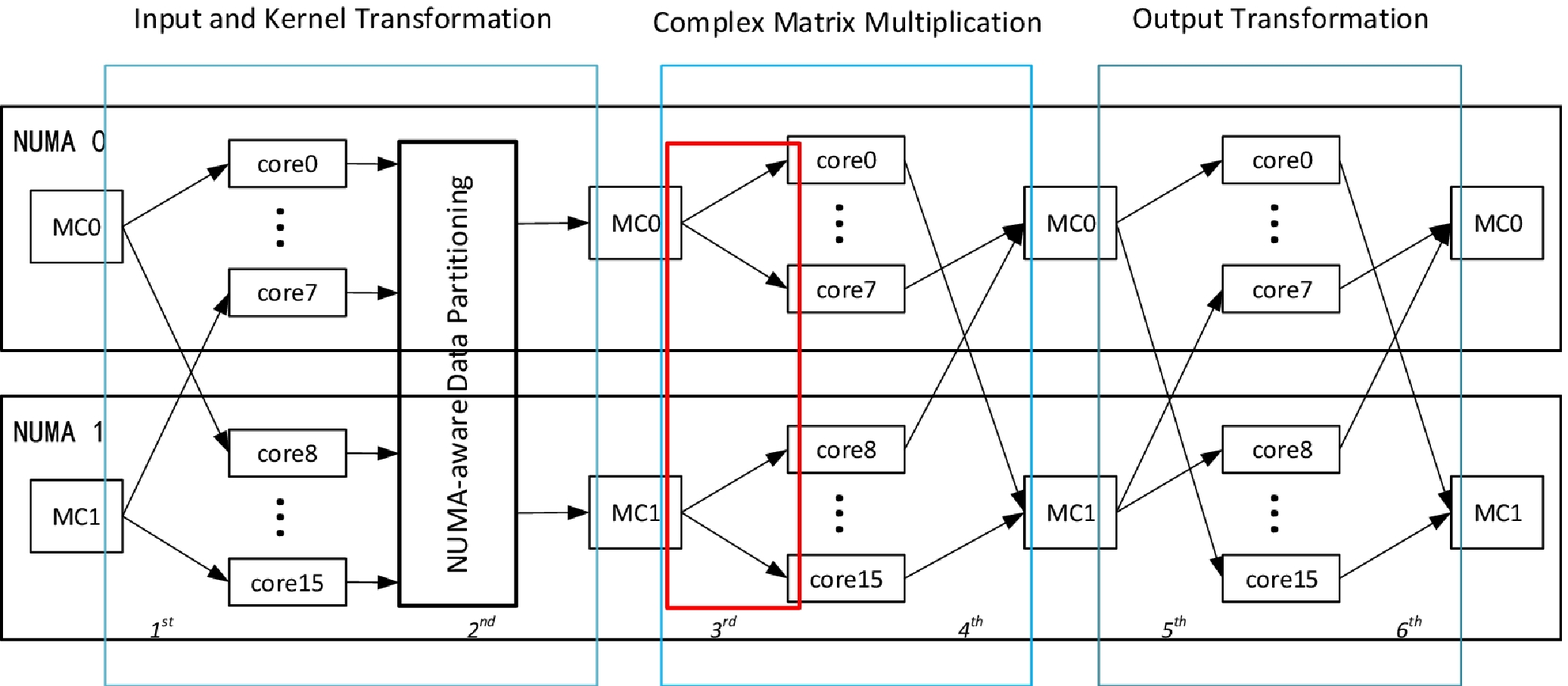}
	\caption{NUMA communication in nFFT, NUMA-aware FFT-based Convolution.}
	\label{fig2:comm:after}
\end{figure*}

According to Section ~\ref{analysis}, a new NUMA-aware algorithm is proposed to reduce remote memory access in FFT-based convolution, as shown in Fig.~\ref{fig2:comm:after}. Compared with wFFT, our nFFT eliminates remote memory access in the fetching ($3^{rd}$ in Fig.~\ref{fig2:comm:after}) of the complex matrix multiplication, by means of rearranging the distribution of the results of the first two transformations on all NUMA nodes and designing the three-level parallelization of the complex matrix multiplication. $P$ complex matrix multiplications are evenly dispatched to all NUMA nodes. 

\subsection{Data Transformation}

\begin{figure}[htb!]
	\centering
	\includegraphics[width=8.5cm, height=5.0cm]{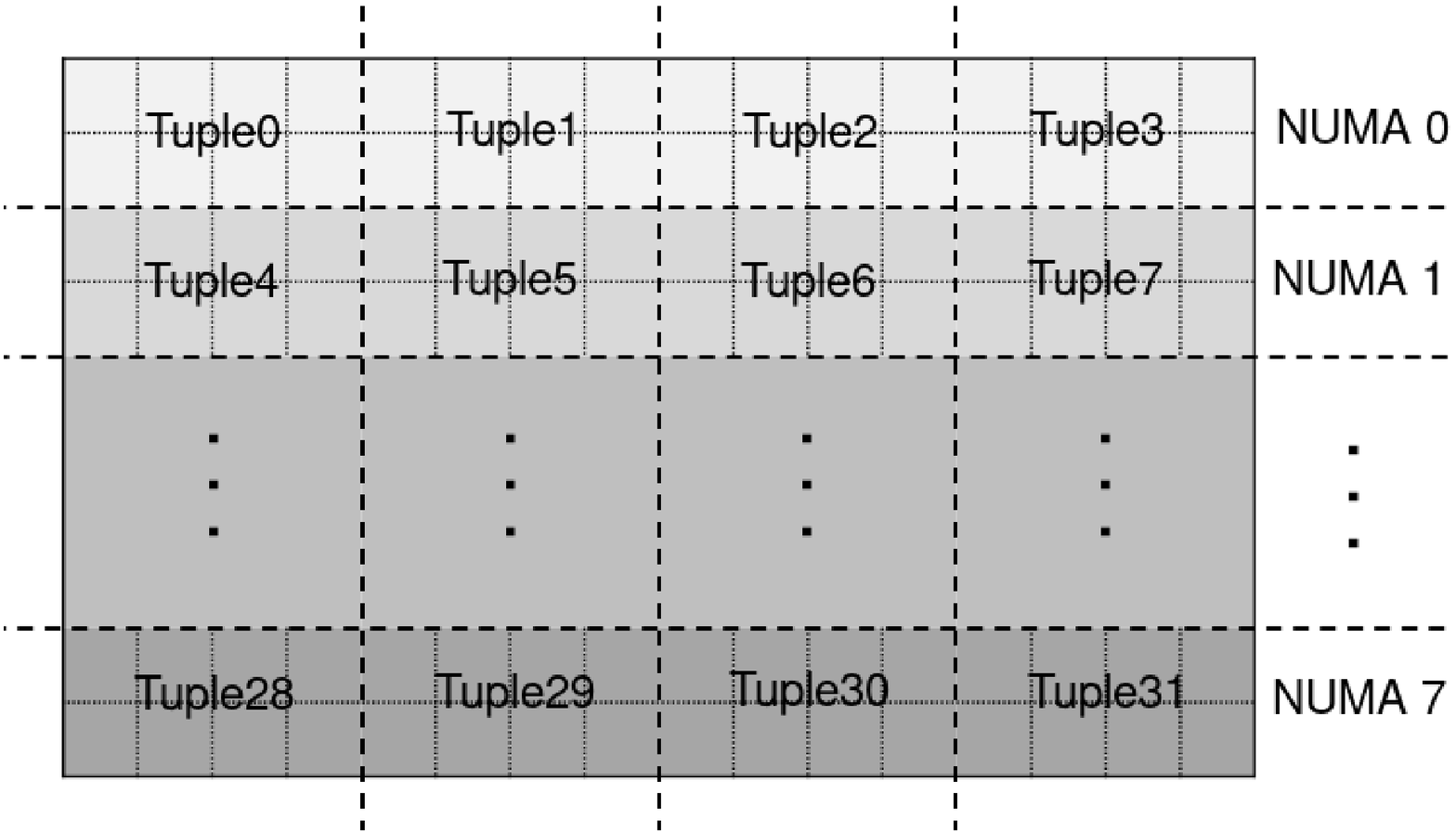}
	\caption{NUMA-aware Tuple Partitioning after transformations.}
	\label{fig3:tuple}
\end{figure}

For the transformations of $I$ and $K$, wFFT only stores half of the FFT results based on the Hermitian symmetry, and the other half are obtained by complex conjugate. For example, if the tile size is $\delta \times \delta$, only $\delta \times \delta$ real numbers is stored. In the following complex matrix multiplication, we still use vector units to deal with $L$ matrix multiplications in parallel. Thus, a transformed tile is further divided into tuples of $2 \times L$ size, and $P$ is equal to $\frac{\delta \times \delta}{2}$. There are $\frac{P}{L}$ tuples in a tile after the transformations. Our nFFT introduce a NUMA-aware tuple partitioning method shown in Fig.~\ref{fig3:tuple}. The method stores the corresponding $\frac{P}{L \times N}$ tuples of $G$ and $D$ on the same nodes, where $N$ represents the number of NUMA nodes. The partitioning in our nFFT just stores the specified tuples on the specified NUMA nodes where the corresponding task resides, while the transformations in wFFT randomly store all tuples on all NUMA nodes. For example, the tile size is $16 \times 16$ and the number of NUMA nodes is 8, so the first and second four tuples are stored back to NUMA 0 and NUMA 1, respectively. Compared with wFFT, nFFT usually doesn't incur huge additional overhead in three other stages except the complex matrix multiplication. For a specified NUMA node, the transformations of $I$ and $K$ also store their results in the access order of the complex matrix multiplication.

For the inverse transformation of $O$, all tuples in a tile are needed to be gathered together. There are two methods to deal with the operation. The one is the complex matrix multiplication stores its results on the specified node and the 5th memory access in Fig.~\ref{fig2:comm:after} fetches the specified tuples from the specified nodes, which is the reverse process of NUMA-aware tuple partitioning. The other is the complex matrix multiplication randomly store its results on all NUMA nodes, and the 5th memory access randomly fetches all the tuples in a tile into a node. According to our experiments, even though the former can usually achieve higher performance in the complex matrix multiplication, the latter usually has better overall performance than the former. As a result, nFFT still adopts the latter method like wFFT.

\subsection{Complex Matrix Multiplication}
As shown in Fig.~\ref{fig2:comm:after}, each node performs its own matrix multiplication, and the detailed algorithm of the complex matrix multiplication is shown in Algorithm~\ref{alg:nfft:cmm}. There are node-level, core-level and vector-level parallelizations. For the node-level parallelization (Line~\ref{node:par}), each NUMA node carried out $P/N$ complex matrix multiplications on its own cores, which fetch the sub-tensors of the tensors $G$ and $D$ by means of the NUMA local memory access. For the core-level parallelization (Lines~\ref{cmm:t1} -~\ref{cmm:t3}), all cores of a specified NUMA node deal with the complex matrix multiplications in parallel. The order of the loops in Lines~\ref{cmm:t0} -~\ref{cmm:t4} are determined by maximizing data resue in two-level caches of Phytium FT-2000plus. For the vector-level parallelization (Line~\ref{cmm:vector}), vector units in a core are used to process $L$ matrix multiplications simultaneously. The detailed derivation of the block sizes ($B_r$, $C'_r$, $C_{l1}$ and $C'_{l2}$) can refer to~\cite{wang2020optimizing}.

\begin{algorithm}[htb!]
	\SetAlgoNoLine
	\SetAlgoNoEnd	
	\DontPrintSemicolon
	\SetKwData{Left}{left}\SetKwData{This}{this}\SetKwData{Up}{up}
	\SetKwFunction{Union}{Union}\SetKwFunction{FindCompress}{FindCompress}
	\SetKwInOut{Input}{input}\SetKwInOut{Output}{output}
	\Input{$G$, $D$}
	\Output{$Z$}
	$N$ is the number of NUMA nodes.\;
	$P=\varphi^2/(2\times L)$, where $\varphi^2$ is the tile size, and $L$ is the vector register width given a specific datatype.\;
	$\rm X \times \Delta$ is the number of tiles in each feature map.\;
	$B_r$, $C'_r$, $C_{l1}$ and $C'_{l2}$ are the block sizes in complex matrix multiplications. \;
	$z'$, $g'_{n}$ and $d'_{n}$ are the sub-tensors of the tensors $Z$, $G$ and $D$, where $g'_{n}$ and $d'_{n}$ are located on the $n$th NUMA node, and $z'$ are distributed on all the NUMA nodes.\;
	\tcp{Node-level Parallelization}
	\ForPar{$\delta = 0 \colon P/N \colon P$\label{node:par}} {
		Get the ID ($n$) of a NUMA node.\;
		\For{$\delta_b =0 \colon 1 \colon P/N $\label{cmm:s}} {
		\For{$cs =0 \colon C_{l1} \colon C$\label{cmm:t0}} {
			\tcp{Core-level Parallelization}
			\ForPar{$cs'=0 \colon C'_{l2} \colon C'$\label{cmm:t1}} {
				\ForPar{$bs=0 \colon B_r \colon B$\label{cmm:t2}} {
					\ForPar{$\mu = 0 \colon 1 \colon \rm X \times \Delta$\label{cmm:t3}} {
						\For{$cofs' =0 \colon C'_{r} \colon C'_{l2}$\label{cmm:t4}} {
							\tcp{Micro-kernel: Vector-level Parallelization}
							$z'[{B_r}][{C'_r}][2 \times L] += \sum\nolimits_{c = cs}^{cs + {C_{l1}}} {{g'_{n,c}}[{C'_r}][2 \times L]}  \times {d'_{n,c}}[{B_r}][2 \times L]$ \label{cmm:vector}\;
							store $z'$ back to matrices $Z$\label{cmm:e}
						}
					}
				}
			}
		}
	}
	}
	\caption{Complex Matrix Multiplication of NUMA-aware FFT-based Convolution.\label{alg:nfft:cmm}}
\end{algorithm}

For the implementation, nFFT uses the function \textit{numa\_alloc\_onnode()} to allocate the memory space on the specified nodes, which is for storing the tensors $G$ and $D$. The left tensors are evenly distributed on all the NUMA nodes. There are $U$ cores in a NUMA node and a total of $N$ NUMA nodes. On Phytium FT-2000plus, each core usually runs one thread for the optimal performance. Thus, a total of $N \times U$ threads are created, and divided into $N$ groups. The function \textit{pthread\_setaffinity\_np()} is called to bind each thread in one group to all $U$ cores of a specified NUMA node. 

\section{Experimental Results}
\label{per_result}

This section introduces the performance comparison of our nFFT against wFFT on Phytium FT-2000plus, and an in-depth analysis of the results.

\subsection{Experimental Setup}
We benchmark 17 unit stride convolutional layers from three popular CNNs: Alexnet\cite{Alexnet2012}, VGG\cite{vgg2014} and Resnet \cite{resnet50he}. The detailed configurations of these convolutional layers is shown in Table~\ref{tab1}, where A, V, and R stand for Alexnet, VGG and Resnet, respectively. Three common batch sizes are set, namely 32, 64 and 128.

\begin{table}
	\caption{The configuration parameters of all tested convolutonal layers}\label{tab1}
	\begin{center}
		\renewcommand{\arraystretch}{1.4}
		\setlength\tabcolsep{3pt}
		\begin{tabular}{lcccccc}
			\toprule[1pt]
			Conv Layers & $B$ & $C$ & $C'$  & $H_i \times W_i$ & $H_k \times W_k$  \\		
			\hline
			Vconv1.1	& 32/64/128 & 3 & 64 & 224 $\times$ 224 & 3 $\times$ 3 \\
			Vconv1.2	& 32/64/128 & 64 & 64 & 224 $\times$ 224 & 3 $\times$ 3 \\	
			Vconv2.1	& 32/64/128 & 64 & 128 & 112 $\times$ 112 & 3 $\times$ 3 \\	
			Vconv2.2	& 32/64/128 & 128 & 128 & 112 $\times$ 112 & 3 $\times$ 3 \\	
			Vconv3.1	& 32/64/128 & 128 & 256 & 56 $\times$ 56 & 3 $\times$ 3 \\	
			Vconv3.2	& 32/64/128 & 256 & 256 & 56 $\times$ 56 & 3 $\times$ 3 \\
			Vconv4.1	& 32/64/128 & 256 & 512 & 28 $\times$ 28 & 3 $\times$ 3 \\
			Vconv4.2	& 32/64/128 & 512 & 512 & 28 $\times$ 28 & 3 $\times$ 3 \\	
			Vconv5	& 32/64/128 & 512 & 512 & 14 $\times$ 14 & 3 $\times$ 3 \\
			\hline
			Aconv2	& 32/64/128 & 48 & 128 & 27 $\times$ 27 & 5 $\times$ 5 \\
			Aconv3	& 32/64/128 & 256 & 384 & 13 $\times$ 13 & 3 $\times$ 3 \\
			Aconv4	& 32/64/128 & 192 & 192 & 13 $\times$ 13 & 3 $\times$ 3 \\
			Aconv5	& 32/64/128 & 192 & 128 & 13 $\times$ 13 & 3 $\times$ 3 \\
			\hline
			Rconv2.2	& 32/64/128 & 64 & 64 & 56 $\times$ 56 & 3 $\times$ 3 \\
			Rconv3.2	& 32/64/128 & 128 & 128 & 28 $\times$ 28 & 3 $\times$ 3 \\
			Rconv4.2	& 32/64/128 & 256 & 256 & 14 $\times$ 14 & 3 $\times$ 3 \\
			Rconv5.2	& 32/64/128 & 512 & 512 & 7 $\times$ 7 & 3 $\times$ 3 \\
			\bottomrule[1pt]
		\end{tabular}
	\end{center}
\end{table}

We measure and analyze the performance of wFFT and NUMA-aware FFT-based convolution (nFFT) on Phytium FT-2000plus. In both wFFT and nFFT, the tile size is $16 \times 16$. All the tests are carried out 100 times, and the median execution time among all the iterations is reported as the final performance of a test.

\subsection{Performance Comparison}

\begin{figure*}[htb!]
	\centering
	\includegraphics[width=15.0cm]{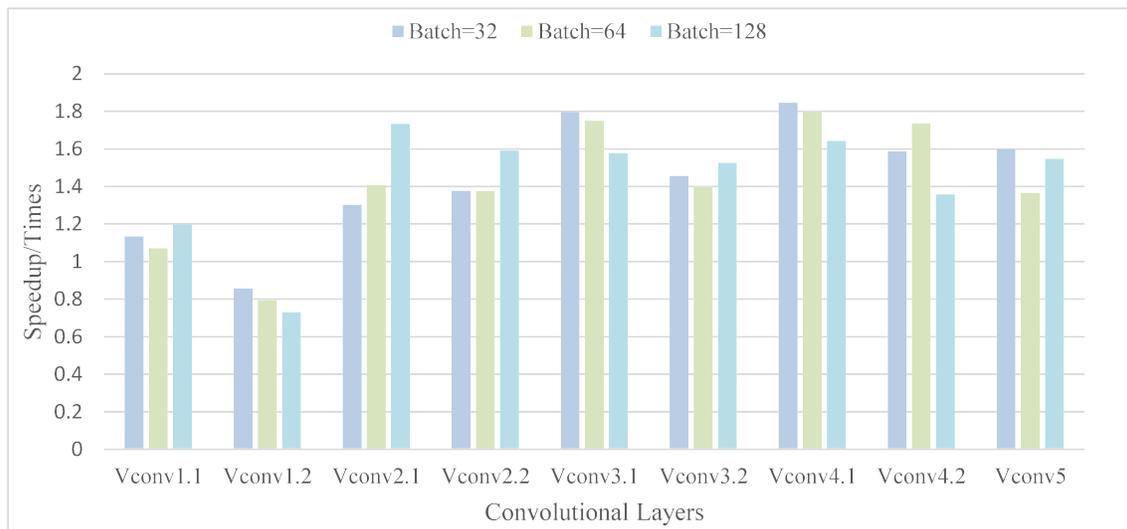}
	\caption{Speedup of our NUMA-aware FFT-based convolution implementation (nFFT) based on wFFT on all 8 NUMA nodes of Phytium FT-2000plus, where the column bars from left to right indicate convolutional layers from VGG.}
	\label{fig4:vgg}
\end{figure*}

\begin{figure*}[htb!]
	\centering
	\includegraphics[width=15.0cm]{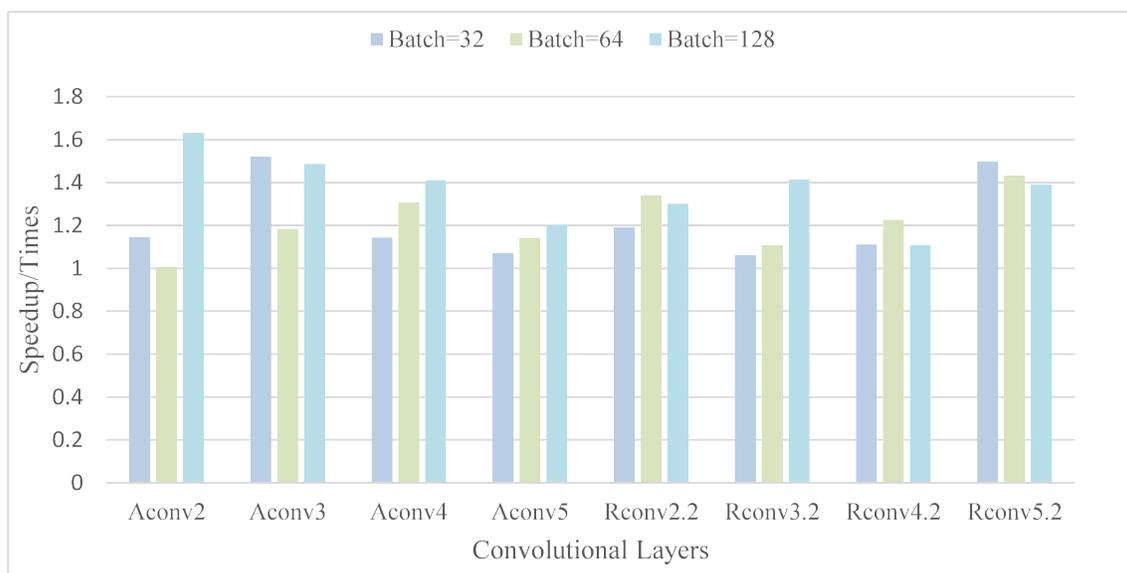}
	\caption{Speedup of our NUMA-aware FFT-based convolution implementation (nFFT) based on wFFT on all 8 NUMA nodes of Phytium FT-2000plus, where the column bars from left to right indicate convolutional layers from Alexnet and Resnet.}
	\label{fig5:resnet-alexnet}
\end{figure*}

The performance comparison between wFFT and our NUMA-aware FFT-based convolution is shown in Fig.~\ref{fig4:vgg} -~\ref{fig5:resnet-alexnet}. In these figures, the abscissa represents different convolutional layers, the ordinate indicates the speedup of nFFT based on wFFT, and the column bars of different colors represent different batch sizes.

Fig.~\ref{fig4:vgg} shows our nFFT is better than wFFT for most of the convolutional layers from VGG. The maximum speedup of 1.84 times is obtained. For some convolutional layers, such as Vconv1.2, nFFT is worse than wFFT. The main reason is that the three-level parallelization of the complex matrix multiplication in our nFFT can not get much better performance than the two-level parallelization of the complex matrix multiplication in wFFT. 

Fig.~\ref{fig5:resnet-alexnet} shows our nFFT surpasses wFFT for all the convolutional layers from Alexnet and Resnet. For Alexnet and Resnet, our nFFT can get the speedups of 1.01 - 1.62 and 1.06 - 1.50 times against wFFT, respectively.

\subsection{Performance Analysis}
To evaluate our optimizations, we further glean some events from  performance monitor units (PMU) on Phytium FT-2000plus. As the PMUs don't support NUMA-related events, we mainly collect two events: Level-2-data-cache-refill and Level-2-data-cache-access. We compare nFFT and wFFT using the Level-2 (L2) cache miss rate, which is the L2 cache miss rate averaged among all the L2 caches. 

\begin{figure*}[htb!]
	\centering
	\includegraphics[width=17.0cm, height=6.0cm]{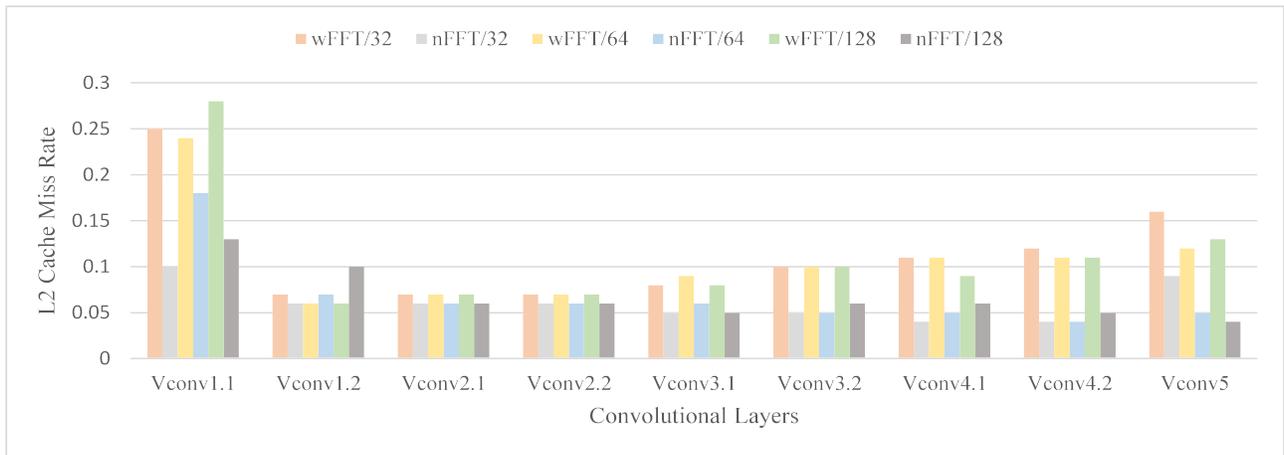}
	\caption{Level-2 cache miss rate comparison between our NUMA-aware FFT-based convolution implementation (nFFT) and wFFT on all 8 NUMA nodes of Phytium FT-2000plus, where the column bars from left to right indicate convolutional layers from VGG.}
	\label{fig6:vgg}
\end{figure*}

\begin{figure*}[htb!]
	\centering
	\includegraphics[width=17.0cm, height=6.0cm]{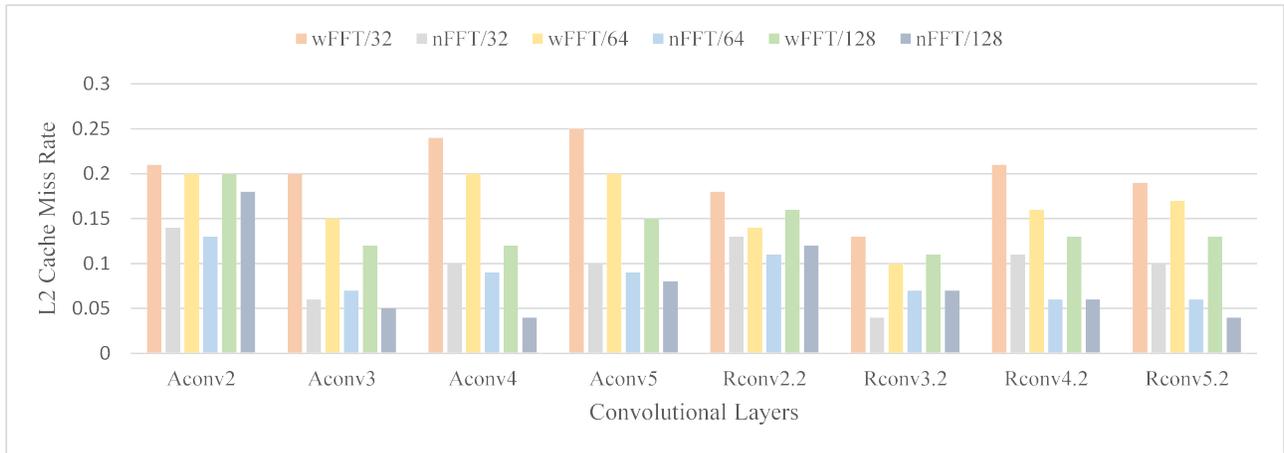}
	\caption{Level-2 cache miss rate comparison between our NUMA-aware FFT-based convolution implementation (nFFT) and wFFT on all 8 NUMA nodes of Phytium FT-2000plus, where the column bars from left to right indicate convolutional layers from Alexnet and Resnet.}
	\label{fig7:resnet:alexnet}
\end{figure*}

Fig.~\ref{fig6:vgg} -~\ref{fig7:resnet:alexnet} show the level-2 cache miss rate comparison between our NUMA-aware FFT-based convolution and wFFT. For the convolutional layers of VGG in Fig.~\ref{fig6:vgg}, our nFFT can get lower miss rate than wFFT in most cases. In Fig.~\ref{fig7:resnet:alexnet}, our nFFT is superior to wFFT on all the convolutional layers of Alexnet and Resnet. At the same time, It is worth nothing that L2 cache misses may have different effects on the performance, depending on whether the misses are caused by local memory access or remote memory access.

\section{Conclusion}
\label{conclusion}
In this paper, we have presented a NUMA-aware FFT-based convolution algorithm on ARMv8 many-core CPUs with NUMA architectures. Based on the analysis of the influence of NUMA architectures on FFT convolution, the algorithm redesigns the parallelization of the complex matrix multiplication, and reorders the results of input and kernel FFT transformations. For the complex matrix multiplication, the three-level parallelization is proposed, including node-level, core-level and vector-level parallelization, and the remote memory access is reduced as much as possible without substantially increasing transformations overhead. The algorithm distributes the results of input and kernel FFT transformations on the specified NUMA nodes, based on the three-level parallel implementation of the complex matrix multiplication. Our NUMA-aware algorithm is evaluated on a ARMv8-based Phytium FT-2000plus 64-core CPU. For the convolutional layers in the popular networks, our nFFT is much better than the prior work in most cases. And the level-2 cache miss rates of our nFFT and the prior work are directly compared.

In the future, we will focus on the optimization of Recurrent Neural Networks on ARMv8 many-core CPUs with NUMA architectures.

\bibliographystyle{IEEEtran}
\bibliography{ref-20210325}

\end{document}